\documentstyle[preprint,aps]{revtex}
\font\tenrm=cmr10

\begin{document}

\preprint{
\font\fortssbx=cmssbx10 scaled \magstep2
\hbox to \hsize{
\includegraphics{uwlogo.ps}
\hskip.5in \raise.1in\hbox{\fortssbx University of Wisconsin - Madison}
\hfill$\vtop{\hbox{\bf MAD/PH/842}
                \hbox{\bf RAL-94-046}
                \hbox{\bf hep-ph@xxx/yymmddd}
                \hbox{July 1994}}$ }
}

\title{\vspace*{.1in}
Constraints on SUSY-GUT unification \\ from $b\to s\gamma$ decay}

\author{V. Barger$^a$,
M.S.~Berger$^a$\footnote{\tenrm Address after Sept. 1, 1994: Physics
Department, Indiana University, Bloomington, IN 47405.},
P.~Ohmann$^a$\footnote{\tenrm Address after Sept. 1, 1994: Deutsches
Elektronen Synchrotron DESY, D-22603 Hamburg, Germany.}
and R.J.N. Phillips$^b$}

\address{
$^a$Physics Department, University of Wisconsin, Madison, WI53706, USA\\
$^b$Rutherford Appleton Laboratory, Chilton, Didcot, Oxon OX11 0QX, UK}

\maketitle

\thispagestyle{empty}

\begin{abstract}
The top-quark Yukawa infrared fixed-point solution of the renormalization
group equations, with minimal supersymmetry and GUT unification, defines a
low-energy spectrum of supersymmetric particle masses in terms of a few
GUT-scale parameters assuming universal boundary conditions.
We give predictions in this model for the inclusive
$b\to s\gamma$ branching fraction and investigate the impact of
non-universal scalar mass boundary conditions. We find our results do not
depend significantly on the value of the GUT-scale trilinear coupling $A^G$.
The small $\tan \beta $ region favors predictions for the inclusive
$b\to s\gamma$ branching fraction close to that of the Standard Model
value. Nevertheless forthcoming experimental results can
eliminate some regions of the GUT parameter space.
\end{abstract}

\newpage

It is now widely appreciated that the $b\to s\gamma $ inclusive branching
fraction is potentially sensitive to new physics beyond the Standard
Model (SM), such as charged Higgs bosons and supersymmetric
particles\cite{bhp,bertolini,bg,etc2}.
The inclusion of QCD corrections to the one-loop amplitudes significantly
enhances the decay rate and these calculations have been steadily
refined\cite{gsw,qcd,qcd2}. The SM amplitude is dominated by the loop with an
intermediate weak boson $W$ and top quark $t$, and the branching fraction
depends on the mass of $t$ for which there is now some direct
evidence\cite{cdftop}. With supersymmetry (SUSY) there are additional
contributions, the most important coming from loops with charged Higgs
$H^{\pm }$ and $t$, charginos $\chi _i^{\pm }$ and stop $\tilde{t}_i$.
The phenomenology can be complex in general, because of the potential for
cancellations between different loop contributions, and because of the large
number of possible parameters. However, considerations of supersymmetric grand
unification\cite{dimop}
(GUT) provide great simplification, with many low-energy
parameters determined by just a few GUT-scale parameters through the
renormalization group equations. In this Letter we investigate the predictions
for $b\to s\gamma$ based on the RGE analysis of Ref.~\cite{bbo}, and show that
measurements of the $B\to X_s\gamma $ inclusive decay rate can exclude
otherwise allowed regions for the GUT parameter space.

An attractive possibility is the realization of an infrared fixed-point
solution\cite{pr} for the top-quark Yukawa coupling $\lambda _t$ in
supersymmetric GUT models\cite{bbo,bbhz,lp,carena}. In this scenario the
top-quark mass $m_t$ is related to the ratio of the two Higgs vacuum
expectation values $v_2/v_1=\tan \beta $ by\cite{bbhz}
\begin{equation}
m_t({\rm pole})\simeq(200\;{\rm GeV})\sin \beta\;.
\end{equation}
In the following discussion we use the value $m_t(pole)=168$ GeV
and $\alpha _s(M_Z^{})=0.120$,
for which the RGE solutions for the sparticle mass spectra
were constructed  in Ref.~\cite{bbo};
this is consistent with the value $m_t=174\pm 10^{+13}_{-12}$
experimentally observed from the top-quark candidate
events\cite{cdftop}.
The universal soft SUSY-breaking parameters at the GUT scale $M_G^{}$ are
$m_0$, $m_{1/2}$, $\mu $, $A$, $B$; here $m_0$ is the scalar mass, $m_{1/2}$
is the gaugino mass, $\mu $ is the Higgs mass mixing, and $A$, $B$ are
trilinear, bilinear scalar couplings respectively. In the ambidextrous
approach of Ref.~\cite{bbo} (see also Ref.~\cite{kelley})
used to solve the RGE integration, $m_{1/2}$,
$m_0$, and $A$ are input at the GUT scale.
Near the fixed point, the value of the trilinear coupling $A_t$ that enters
into squark mixing is drawn to a fixed value (for a given
$\alpha_3(M_Z^{})$)\cite{carena,drees},
so that the dependence of the rate on the initial value
$A^G$ is minimal.
These soft SUSY-breaking parameters are evolved
down to the electroweak scale $M_Z^{}$, and the values $|\mu (M_Z^{})|$ and
$B(M_Z)$ are determined from the requirement that the electroweak symmetry be
broken radiatively. By this procedure, all sparticle masses and couplings are
determined in terms of GUT parameters $m_0$ and $m_{1/2}$ for a specified
sign of $\mu $. Thus we can predict the $b\to s\gamma $ decay rate over the
region of $(m_0,\;m_{1/2})$ parameter space allowed by other constraints,
which are experimental lower bounds on sparticle masses and a naturalness
criterion taken to be $|\mu |< 500$ GeV in Ref.~\cite{bbo}.

For calculating the $b\to s\gamma $ decay rate, we use the latest analysis of
QCD corrections by Buras et al.\cite{qcd}, which evaluates the full
$8\times 8$ anomalous dimension matrix. We do not
take into account the small effect
of running the scale of the decay process from $Q=m_t$ down to $M_W$
(estimated in Ref.~\cite{cg}), beyond the usual effect of
running from $Q=M_W$ down to $m_b$. The QCD effects also receive
theoretical corrections from other large mass splittings in the supersymmetric
spectrum\cite{anlauf}.

The ratio of
$\Gamma (b \rightarrow s \gamma)$ to
the inclusive semileptonic decay width is then given by
\begin{equation}
{{\Gamma(b\rightarrow s \gamma)}\over {\Gamma (b\rightarrow ce\nu )}}
={{6\alpha }\over {\pi \rho \lambda }}
{{|V_{ts}^*V_{tb}|^2}\over {|V_{cb}|^2}}|c_7(m_b)|^2 \;, \label{ratio}
\end{equation}
where $\alpha $ is the electromagnetic coupling.   The phase-space factor
$\rho $ and the QCD correction factor $\lambda $ for the semileptonic process
are given by $\rho = 1 - 8r^2 + 8r^6 - r^8 - 24r^4\ln(r)$ with $r= m_c/
m_b=0.316\pm 0.013$  and
$\lambda = 1 - {2\over 3} f(r,0,0)\alpha_s^{}(m_b)/\pi $
with $f(r,0,0)=2.41$\cite{QCDcorr}.

The formulas for the renormalization group coefficient in Eq.~(\ref{ratio})
are\cite{bg,88}
\begin{eqnarray}
c_7(m_b)=\left [{{\alpha _s(M_W)}\over {\alpha _s(m_b)}}\right ]^{16/23}
\Bigg \{c_7(M_W) &-& {8\over 3} c_8(M_W)\left [1-\left (
{{\alpha _s(m_b)}\over {\alpha _s(M_W)}}\right )^{2/23}\right ]\Bigg \}
\nonumber \\
&+&\sum _{i=1}^8 h_i\left ({{\alpha _s(M_W)}\over {\alpha _s(m_b)}}
\right )^{\alpha _i}
\;, \label{c7} \\
   c_7(M_W) = {3\over 2} xf_{\gamma}^{(1)}(x) +{1\over 2}y f_{\gamma}^{(2)}(y)
&+&
{1\over {2\tan^2 \beta}} yf_{\gamma}^{(1)}(y)
+c_7^{\chi^\pm}
\;, \label{c7II} \\
   c_8(M_W) = {3\over 2} xf_g^{(1)}(x) +{1\over 2}y f_g^{(2)}(y) &+&
{1\over {2\tan^2 \beta}} yf_g^{(1)}(y)
+c_8^{\chi^\pm}
\;, \label{c8II}
\end{eqnarray}
with  $x = (m_t/M_W)^2$, $y = (m_t/m_{H^{\pm}})^2$, and
\begin{eqnarray}
c_7^{\chi^\pm}&=&\sum_{j=1}^2\Bigg\{{{M_W^2}\over {\tilde{m}_{\chi_j}}^2}
\Bigg [|V_{j1}|^2f_{\gamma}^{(1)}(z_1)-\sum_{k=1}^2\Bigg
|V_{j1}T_{k1}-V_{j2}T_{k2}{{m_t}
\over {\sqrt{2}M_W\sin \beta}}\Bigg |^2f_{\gamma}^{(1)}(z_2)\Bigg ]\nonumber \\
&-&{{U_{j2}}\over {\sqrt{2}\cos\beta}}{{M_W}\over {\tilde{m}_{\chi_j}}}\Bigg [
V_{j1}f_{\gamma}^{(3)}(z_1)-\sum_{k=1}^2\Bigg (V_{j1}T_{k1}-V_{j2}T_{k2}{{m_t}
\over {\sqrt{2}M_W\sin \beta}}\Bigg )T_{k1}f_{\gamma}^{(3)}(z_2)\Bigg ]\Bigg
\}\;,\\
c_8^{\chi^\pm}&=&\sum_{j=1}^2\Bigg\{{{M_W^2}\over {\tilde{m}_{\chi_j}}^2}
\Bigg [|V_{j1}|^2f_g^{(1)}(z_1)-\sum_{k=1}^2\Bigg
|V_{j1}T_{k1}-V_{j2}T_{k2}{{m_t}
\over {\sqrt{2}M_W\sin \beta}}\Bigg |^2f_g^{(1)}(z_2)\Bigg ]\nonumber \\
&-&{{U_{j2}}\over {\sqrt{2}\cos\beta}}{{M_W}\over {\tilde{m}_{\chi_j}}}\Bigg [
V_{j1}f_g^{(3)}(z_1)-\sum_{k=1}^2\Bigg (V_{j1}T_{k1}-V_{j2}T_{k2}{{m_t}
\over {\sqrt{2}M_W\sin \beta}}\Bigg )T_{k1}f_g^{(3)}(z_2)\Bigg ]\Bigg \}\;,
\end{eqnarray}
with $z_1=(\tilde{m}/\tilde{m}_{\chi_j})^2$ and
$z_2=(\tilde{m}_{t_k}/\tilde{m}_{\chi_j})^2$ and the matrices $T$, $U$ and $V$
diagonalize the top squark and chargino mass matrices
(see Refs.~\cite{bg,bbo}). The parameter $\tilde{m}$ is the mass of the
(degenerate) squarks from the first two generations.

The coefficients from the $8\times 8$ anomalous dimension matrix are\cite{qcd}
\begin{equation}
\begin{array}{ccccccccc}
\vspace{0.2cm}
a_i = ( &    \frac{14}{23},     &     \frac{16}{23},   &
\frac{6}{23},&-\frac{12}{23},
        &      0.4086,       &      -0.4230,     & -0.8994,  &   0.1456     )
\\
h_i = ( &\frac{626126}{272277}, &-\frac{56281}{51730}, &-\frac{3}{7}, &
-\frac{1}{14},
        &     -0.6494,       &      -0.0380,     & -0.0186,  &  -0.0057     )
\end{array}
\end{equation}
The formulas for the loop contributions $f_{{\gamma},g}$
are given in Ref.~\cite{bg}.

The neutralino and gluino loop contributions have not been included in the
calculation of the $b\to s\gamma $ inclusive rate. However, these
contributions are known to be
small, especially in the low $\tan \beta $ regime.

Our results are displayed in Figs.~\ref{fig:fig2}-\ref{fig:fig4}.
Figure \ref{fig:fig2} and Figure \ref{fig:fig3} show the predicted
value of the inclusive branching fraction $B(b\to s\gamma)$, that is
equal in the spectator approximation to the physical meson branching
fraction $B(B\to X_s\gamma)$; the upper part of these figures
gives the branching fraction
versus $m_0$ for fixed $m_{1/2}$ , while the lower part gives
curves versus $m_{1/2}$ for fixed $m_0$ .  The same
information is displayed in Figure~\ref{fig:fig4}, in the form of contours of
constant $B(b\to s\gamma)$ in the ($m_{1/2},m_0$) parameter plane.
The region is bounded by experimental limits on
the chargino mass $m_{\chi_1^{\pm}}$ and
the lighter stop mass $m_{\tilde t_1}$ and the lighter higgs mass $m_h$
together with the naturalness boundary $|\mu |=500$ GeV.
The shaded region is allowed by these constraints.

It has come to light in recent months that the leading order QCD
calculations are somewhat inadequate for comparison with the current
experimental data;
hence a brief discussion of the theoretical uncertainties in warranted.
The $(m_b)^5$ dependence of the rate for $b\to s \gamma$ is customarily
removed by taking the ratio with semileptonic decays. This, however, does not
entirely eliminate the theoretical uncertainty due to $m_b$ because
the $b$-mass still enters
into the phase space factor which depends on the ratio $m_b/m_c$.
The ratio of CKM matrix entries is taken to be the central value\cite{qcd}
\begin{eqnarray}
{{|V_{ts}^*V_{tb}^{}|}\over |V_{cb}^{}|^2}&=&0.95\pm 0.04\;.
\end{eqnarray}
There is even more substantial uncertainty in the higher order (three-loop)
contributions.
It has been argued\cite{qcd} that the unknown next-to-leading QCD corrections
yield a $\pm 25$\% uncertainty in the branching fraction. This number is
arrived at by varying the unphysical scale $\mu $ from $m_b/2$ to $2m_b$, and
is particularly large since the process $b\to s\gamma $ is dominated by the
QCD corrections. The calculation of the next-to-leading corrections is a
formidable task, and progress in reducing this theoretical uncertainty is
not anticipated in the near future. Furthermore, the value of
$\alpha _s(M_Z^{})$ is not known very precisely, and for $\alpha _s(M_Z^{})$ in
the range $0.12\pm 0.01$, the $b\to s \gamma $ rate changes by roughly 10\%.
Finally, the experimental error in the
measurement of the semileptonic
decay rate must be considered.

Most analyses of the $b\to s \gamma $ process have focused on the
case where the scalar masses are universal at the GUT scale. It has recently
become fashionable to consider situations where the scalar masses take a more
general form but still satisfy the stringent bounds on the flavor changing
neutral currents of the first two generations. This nonuniversality might be
generic\cite{nonuniv},
or it might arise from the running of the scalar masses from a
universal value near the Planck scale to the GUT scale\cite{pp}.
When the gauge group is broken to one of lesser rank
and non-universality in the
soft-supersymmetry breaking terms exists, then there are
in general additional D-term contributions to the scalar
masses\cite{fhkn,kmy}.
One must be careful to include the often-neglected contributions to the
renormalization group equations of the soft-supersymmetry breaking parameters
that arise when non-universality is assumed\cite{faulk}.
The combination
\begin{eqnarray}
{\cal S}&=&
{m^2_{H_2}}-{m^2_{H_1}}+ {\bf Tr} [{\bf M}^2_{Q_L}-{\bf M}^2_{L_L}
-2{\bf M}^2_{U_R}
+{\bf M}^2_{D_R}+{\bf M}^2_{E_R}]\;,
\end{eqnarray}
satisfies the (one-loop) scaling equation\cite{kmy}
\begin{eqnarray}
{{d{\cal S}}\over {dt}}&=&
{{2b_1g_1^2}\over {16\pi ^2}}{\cal S}\;,
\end{eqnarray}
so that if it is zero at some scale, for example the
GUT scale, then it is zero for all scales. The renormalization group
coefficient $b_1=33/5$ in the minimal supersymmetric model.
We have used the two-loop renormalization group equations for
which three groups are now in complete agreement\cite{twoloop}.
The supersymmetric spectrum that results will be different from that
obtained with universal masses at the GUT scale.
In a particular example
based on the minimal $SU(5)$ GUT investigated recently
by Polonsky and Pomarol\cite{pp},
each of the sfermion ({\bf 10} and {\bf 5})
and the Higgs masses is separated at the GUT scale due to the evolution
from the Planck scale.
If one takes the sfermion mass to be twice that of the Higgs scalar mass at
the GUT scale, the approach of $B(b\to s\gamma )$
to the Standard Model value is faster than in the
universal case as one increases the scalar masses: see Figures 4 and
5. This results from the contributions of
the comparatively heavier top squarks which enter into
the chargino loop diagrams.

One can also consider the nonuniversality of the two Higgs doublet mass
parameters\cite{mn}.
In this case one finds that the value of $\mu $ needs to be adjusted to achieve
the correct electroweak symmetry breaking as described
by the (tree-level) expression
\begin{eqnarray}
{1\over 2}M_Z^2&=&{{m_{H_1}^2-m_{H_2}^2\tan ^2\beta }
\over {\tan ^2\beta -1}}-\mu ^2 \;. \label{treemin1}
\end{eqnarray}
The charged-Higgs mass can be increased or decreased, but since $\mu $ is large
for small $\tan \beta$, the mass is guaranteed to be large as well.
Consequently the
impact on the $b\to s\gamma $ rate is small. In some small regions of
parameter space, the mixing between the top squarks can be enhanced.
However the overall qualitative behaviour depicted in
Figures 1 and 2 is not changed substantially.

The effects of the nonuniversality of the symmetry breaking terms
on $B(b\to s\gamma )$ in the low $\tan \beta $ region can be
summarized as follows: (1) There is little
dependence on the chargino mass and couplings
in the low-$\tan \beta $ fixed-point region since $\mu $ is large
and dominates the chargino mass matrix. Therefore the mass of the
lightest chargino
is approximately $M_2={{\alpha_2}\over {\alpha _G}}m_{1/2}$,
and its coupling is predominantly gaugino.
(2) Consequently, the dominant effect comes from the top squark masses and
mixing. Typically increasing $m_0^{\tilde{f}}$ relative to $m_0^H$
increases the top squark mass and pushes the $b\to s\gamma $ rate closer to
the Standard Model value.
(3) The dependence on $A^G$ is minimal near the top Yukawa coupling fixed
point because the low energy value $A_t$ is driven to a fixed
value\cite{carena,drees}. It has
pronounced effects only in regions where
there is large top squark mixing. This results in the lightest top squark
mass being suppressed (below the experimental limit and one is
even in danger of the mass-squared going
negative) and an enhancement of the chargino contribution to the
$b\to s\gamma $ inclusive rate.

To summarize, we have evaluated the inclusive decay branching fraction
$B(b\to s\gamma)$ through the allowed regions of SUSY-GUT parameters,
for solutions with small $\tan\beta$ near the $\lambda_t$ fixed point,
following Ref.\cite{bbo}.  Our results show that\\
(a) In the low $\tan \beta $ fixed point region, the inclusive rate
for $b\to s\gamma $ is typically close to that
expected in the Standard Model. It
is larger (smaller) than the Standard Model value for
$\mu >0$ ($\mu <0$) when the difference is substantial.\\
(b) An accurate measurement of $B(b\to s\gamma)$ must be coupled with reduced
theoretical uncertainties in order to
constrain the parameters $m_{1/2}$ and $m_0$, or to test the
universality of the soft-supersymmetry breaking parameters.\\
(c) The contours in Figure 3 illustrate the kind of constraints that may be
expected.

\acknowledgements

We thank G.~Park for discussions. This research was supported
in part by the University of Wisconsin Research Committee with funds granted by
the Wisconsin Alumni Research Foundation, in part by the U.S.~Department of
Energy under Contract No.~DE-AC02-76ER00881, and in part by the Texas National
Laboratory Research Commission under Grant Nos.~RGFY93-221 and FCFY9302.
MSB was supported in part by an SSC Fellowship. PO was supported in
part by an NSF Graduate Fellowship.

\newpage
\section*{Figures}

\begin{enumerate}

\item{Predicted values of $B(b\to s\gamma)$ (a) versus $m_0$ for
fixed values of $m_{1/2}$ and (b) versus $m_{1/2}$  for fixed values of
$m_0$ , in the low $\tan \beta $
SUSY-GUT solutions of Ref.~\cite{bbo} with $\mu > 0$. The curves are shown in
intervals of $10$ GeV in $m_0$ and $m_{1/2}$; each curve in (a) corresponds to
a vertical line in (b).
\label{fig:fig2}}

\item{Predicted values of $B(b\to s\gamma)$ (a) versus $m_0$ for
fixed values of $m_{1/2}$ and (b) versus $m_{1/2}$  for fixed values of
$m_0$ , in the low $\tan \beta $
SUSY-GUT solutions of Ref.~\cite{bbo} with $\mu < 0$. The cross indicates
the point at which the lighter top squark mass-squared becomes negative.
\label{fig:fig3}}

\item{Contours of fixed $B(b\to s\gamma)$ in the ($m_{1/2},m_0$)
parameter plane, for (a) $\mu > 0$ and (b) $\mu < 0$. The shaded
areas are allowed by direct experimental and naturalness constraints
\cite{bbo}.\label{fig:fig4}}

\item{Predicted values of $B(b\to s\gamma)$ for $\mu > 0$
(a) versus $m_0^{\tilde{f}}=2m_0^H$
for fixed values of $m_{1/2}$ and (b) versus $m_{1/2}$  for fixed values of
$m_0^H$.
\label{fig:fig5}}

\item{Predicted values of $B(b\to s\gamma)$ for $\mu < 0$
(a) versus $m_0^{\tilde{f}}=2m_0^H$
for fixed values of $m_{1/2}$ and (b) versus $m_{1/2}$  for fixed values of
$m_0^H$.
\label{fig:fig6}}

\end{enumerate}

\end{document}